\documentclass[aps,prd,preprint,groupedaddress]{revtex4-1}

\usepackage{amsmath}
\usepackage{graphicx}
\usepackage{amsbsy}
\usepackage{amssymb}
\usepackage{palatino,avant,graphicx,color}
\usepackage{dcolumn}
\usepackage{bm}
\begin{document}

\title{Flow Patterning in Hele-Shaw Configurations using Non-Uniform Electroosmotic Slip}

\author{Evgeniy Boyko$^\S$}
\author{Shimon Rubin$^\S$}
\author{Amir Gat$^*$}
\author{Moran Bercovici$^*$}
\affiliation{$^\S$Equal contribution\\*Corresponding authors\\Faculty of Mechanical Engineering, Technion - Israel Institute of Technology, Haifa, Israel}

\begin{abstract}
We present an analytical study of electroosmotic flow in a Hele-Shaw
configuration with non-uniform zeta potential distribution. Applying
the lubrication approximation and assuming thin electric double layer,
we obtain a pair of uncoupled Poisson equations for the pressure and
depth-averaged stream function, and show that  the inhomogeneous parts in these equations
are governed by gradients in zeta potential parallel and perpendicular
to the applied electric field, respectively. We obtain a solution
for the case of a disk-shaped region with uniform zeta potential
and show that the flow field created is an exact dipole, even in the
immediate vicinity of the disk. In addition, we study the inverse problem where the desired
flow field is known and solve for the zeta potential distribution
required in order to establish it. Finally, we demonstrate that such
inverse problem solutions can be used to create directional flows
confined within narrow regions, without physical walls. Such solutions
are equivalent to flow within channels and we show that these can
be assembled to create complex microfluidic networks, composed of
intersecting channels and turns, which are basic building blocks in
microfluidic devices. 
\end{abstract}
\maketitle

\section{Introduction}

Electroosmotic flow (EOF) is the motion of a liquid due to interaction
of an externally applied electric field with the net charge in the
diffuse part of an electrical double layer. For solid surfaces
with uniform zeta potential and channel geometry, EOF is characterized
by a uniform plug-like velocity profile. However, in practice, most
surfaces are non-homogeneously charged to some extent, either due
to manufacturing limitations or by intended design (see \cite{Herr} and  \cite{Stone} and references therein). 

EOF in capillaries with non-homogeneous zeta potential distribution
has been studied thoroughly. Anderson and Idol \cite{AndersonIdol} found an exact
solution to EOF through a capillary with axially varying zeta potential,
assuming negligible inertia and a thin Debye layer. Herr \textit{et
al.} \cite{Herr} investigated, analytically and experimentally, EOF through
a cylindrical capillary with an axial step change in zeta potential distribution.
The authors measured experimentally the flow profile and showed good agreement with theoretical predictions.
Ghosal \cite{Ghosal} applied a lubrication approximation to investigate
EOF in an infinitely long channel with slow axial variation in cross-section
geometry and zeta potential. Ghosal showed that
the flow rate through any section can be related to a uniform cylindrical
capillary with an equivalent radius and zeta potential, which
are determined solely by the geometry and surface charge distribution
in the channel. 

Numerous works have also considered EOF
between parallel plates. Ajdari \cite{Ajdari95,Ajdari96} was the first to analytically
study and provide a closed form solution to the two dimensional problem
of EOF between an undulating plate and a flat plate with a periodic
surface charge distribution. Ajdari focused his analysis on the cross-section
of the flow cell and demonstrated that the interaction of periodic
deformations of the plate with periodic distribution of zeta potential
gives rise to net flow generation between the plates, even though
the plates are on average electro-neutral. Focusing on flow between
parallel plates, Long \textit{et al. }\cite{Long} provided an analytical
solution for the three dimensional EOF field due to arbitrary distribution
of zeta potential. In particular, by taking the limit of a small
gap between parallel plates, the authors obtained a solution
associated with a Hele-Shaw case, deducing that a localized defect
in zeta potential distribution induces long-range flow perturbations.
Ajdari \cite{Ajdari01} considered a three dimensional geometry, consisting
of two plates with a spatially varying gap and periodic zeta potential
in one direction. By applying the lubrication theory, Ajdari formulated
the Onsager matrix, which relates the applied pressure gradient and
electric field to flow rate and electric current. Stroock \textit{et
al.} \cite{Stroock} performed an experimental study of EOF in flat shallow
micro-channels having non-uniform zeta potential and compared
the results with those of Ajdari \cite{Ajdari95,Ajdari96} and Long \textit{et
al.} \cite{Long}. Stroock \textit{et al.} considered alternate positive
and negative zeta potential patterns and demonstrated that various
flow types such as multi-directional or circular flow can be generated,
depending on whether the applied field is parallel or perpendicular
to gradient of zeta potential. 

In this work, we aim to study the potential use of such non-uniform
electroomotic flows as a mechanism to create complex microfluidic channel networks,
not requiring solid walls. In sections II and III, we define the problem
and derive a pair of uncoupled Poisson equations for the pressure
and depth-averaged stream function, governed by gradients in zeta potential parallel
and perpendicular to the applied electric field, respectively. In
section IV, we focus on axially symmetric zeta potential distributions,
and show that the flow field created is an exact dipole in the immediate
vicinity of a disk with uniform zeta potential. In section V,
we show the effect of the orientation (relative to the electric field)
of thin lines with uniform zeta potential on the pressure and
vorticity fields. In section VI, we study the inverse problem where
the desired flow field is known and solve for the zeta potential
distribution required in order to establish it. We then demonstrate
that such inverse problem solutions can be used to create flows of desired directionality confined within narrow regions, without physical walls. We show
that these can be assembled to create complex microfluidic networks,
and provide a specific example of Y-junction leading to a meandering
channel. 

\section{Problem Definition}

We here denote dimensional variables by tildes and normalized variables
without tildes. Figure 1 presents a schematic illustration of the geometry and the relevant physical quantities. We consider creeping flow in a narrow gap
between two parallel plates subjected to a uniform in-plane electrostatic
field $\tilde{\mathit{\mathrm{\boldsymbol{\mathit{E}}}}}_{||}$. We
employ a Cartesian coordinate system $(\tilde{x},\tilde{y},\tilde{z})$
whose $\tilde{x}$ and $\tilde{y}$ axes lie at the lower plane and
$\tilde{z}$ is perpendicular thereto. The gap between the lower and upper plate is $\tilde{h}$. Each has an arbitrary zeta potential distribution, respectively defined as $\tilde{\zeta}^{L}(\tilde{x},\tilde{y})$ and $\tilde{\zeta}^{U}(\tilde{x},\tilde{y})$, which vary over a characteristic length scale $\tilde{l}$ in the $\tilde{x}-\tilde{y}$ plane.
 Hereafter, we adopt the $\Vert$ and $\bot$ subscripts to denote parallel and
perpendicular vector components to the $\tilde{x}-\tilde{y}$ plane, respectively. 
\begin{figure}[h]
\centering{}\includegraphics{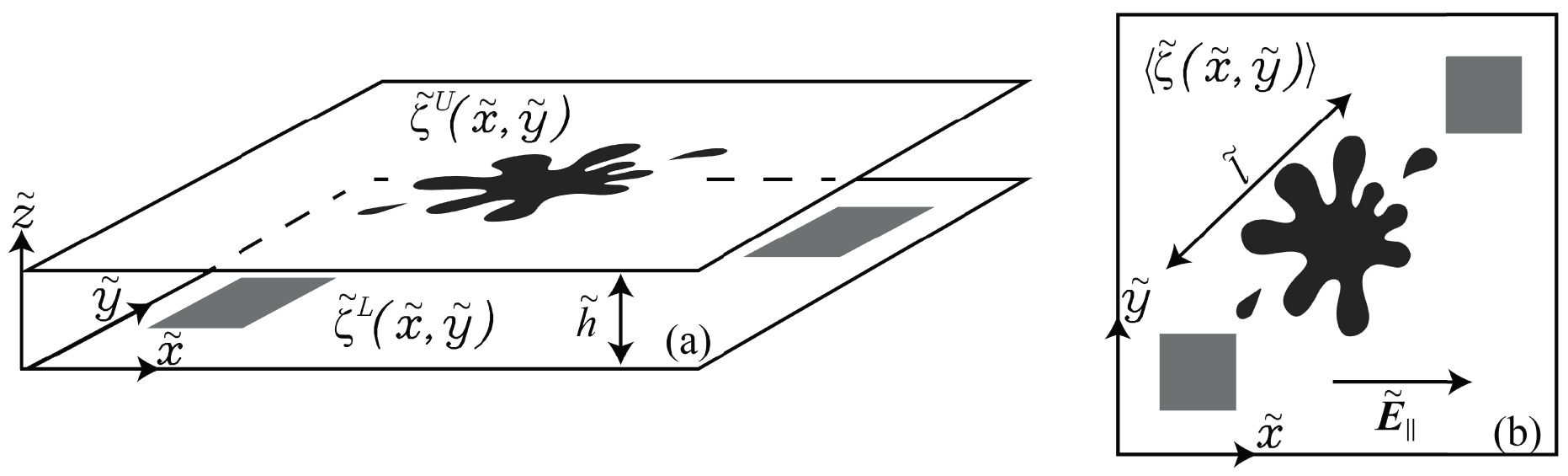}
\protect\caption{\textit{Schematic illustration of the problem. (a) We consider a configuration
	consisting of two parallel plates, separated by a gap $\tilde{h}$.
	Each of the plates is functionalized with arbitrary zeta potential
	distribution, $\tilde{\zeta}^{L}(\tilde{x},\tilde{y}),\,\tilde{\zeta}^{U}(\tilde{x},\tilde{y}),$
	on the lower and upper plates, respectively. The fluid is subject to a uniform electric field
 	$\tilde{\mathit{\mathrm{\boldsymbol{\mathit{E}}}}}_{\Vert}$.
	(b) Averaging over the depth of the cell yields a 2D problem, in which
	$\left\langle \tilde{\zeta}\right\rangle $ denotes the average potential
	of the two plates, $\left\langle \tilde{\zeta}\right\rangle =(\tilde{\zeta}^{U}+\tilde{\zeta}^{L})/2.$ }}
\end{figure}
\vspace{-0.2in}

The continuity equation for an incompressible fluid is 
\begin{equation}
\tilde{\boldsymbol{\nabla}}\cdot\mathbf{\tilde{\boldsymbol{\mathit{u}}}=\mathrm{0}}.\label{eq:Continuity}
\end{equation}
Assuming a thin electric double layer regime (see \cite{Ajdari95,Ajdari96}), the momentum equation in the bulk is given by 
\begin{equation}
	\tilde{\rho} \left( \frac{\partial\tilde{\mathit{\mathbf{\boldsymbol{\mathit{u}}}}}}	
          {\partial\tilde{t}}+\boldsymbol{\tilde{u}}\cdot\tilde{\boldsymbol{\nabla}}\boldsymbol{\tilde{u}} \right)=
          -\tilde{\boldsymbol{\nabla}}\mathit{\tilde{p}}+\tilde{\eta}\tilde{\nabla}^{\mathrm{2}}\tilde{\boldsymbol{u}}.
\label{eq:Momentum}
\end{equation}
where $\boldsymbol{\tilde{u}}=(\tilde{u},\tilde{v},\tilde{w})$
is the velocity vector, $\tilde{t}$ is time and $\tilde{\rho},\,\tilde{\eta}$
and $\tilde{p}$ denote the fluid density, dynamic viscosity and pressure,
respectively. We account for the body forces acting on the double
layer by using the Helmholtz-Smoluchowski slip boundary conditions, \cite{Hunter}
\begin{equation}
\tilde{\boldsymbol{u}}_{\Vert}|_{\tilde{z}=0}=-\frac{\tilde{\varepsilon}\tilde{\zeta}^{L}\tilde{\boldsymbol{E}}_{\Vert}}{\tilde{\eta}},\quad\tilde{\boldsymbol{u}}_{\Vert}|_{\tilde{z}=\tilde{h}}=-\frac{\tilde{\varepsilon}\tilde{\zeta}^{U}\tilde{\boldsymbol{E}}_{\Vert}}{\tilde{\eta}},\quad\tilde{\boldsymbol{u}}_{\bot}|_{\tilde{z}=0,\,\tilde{h}}=0,\label{eq:Boundary-conditions}
\end{equation}
where $\boldsymbol{\tilde{u}}_{\Vert}=(\tilde{u},\tilde{v})$, $\boldsymbol{\tilde{u}}_{\bot}=\tilde{w}\boldsymbol{\hat{z}}$
and $\tilde{\varepsilon}$ is the fluid permittivity, and $\tilde{\zeta}^{U},\,\tilde{\zeta}^{L}$
are zeta potential distribution on the upper and lower plates,
respectively. 

\section{Governing Equations for EOF in a Hele-Shaw Cell with Non-Uniform Zeta Potential}

Hereafter we denote characteristic scales of the system by an asterisk
superscript. The typical magnitude of the flow velocity in the $\tilde{x}-\tilde{y}$
plane, $\tilde{u}^{*}$, is determined by the Helmholtz-Smoluchowski
slip condition as $\tilde{u}^{*}=-\tilde{\varepsilon}\tilde{\zeta}^{*}\tilde{E}{}^{*}/\tilde{\eta}$,
where $\tilde{\zeta}^{*}$ is characteristic value of zeta potential
and $\tilde{E}{}^{*}$ is characteristic externally applied electric
field. The characteristic time scale is $t^{*}=\tilde{l}/\tilde{u}^{*}$,
whereas the characteristic velocity in the $\boldsymbol{\hat{z}}$
direction, $\tilde{w}^{*}$, and the characteristic pressure, $\tilde{p}^{*}$,
remain to be determined from scaling arguments. 

We introduce the following normalized quantities, $(x,y,z)=(\tilde{x}/\tilde{l},\tilde{y}/\tilde{l},\tilde{z}/\tilde{h})$, $t=\tilde{t}/\tilde{t}^{*},$ $(u,v,w)=(\tilde{u}/\tilde{u}^{*},\tilde{v}/\tilde{u}^{*},\tilde{w}/\tilde{w}^{*})$, $p=\tilde{p}/\tilde{p}^{*},$ $\zeta=\tilde{\zeta}/\tilde{\zeta}^{*}$ and $\boldsymbol{E_{\Vert}}=\tilde{\boldsymbol{E}}_{\Vert}/\tilde{E}{}^{*}.$
We restrict our analysis to a shallow flow chamber,
\begin{equation}
\epsilon=\frac{\tilde{h}}{\tilde{l}}\ll1,\label{eq:Epsilon}
\end{equation}
and a negligible inertia regime. The latter is characterized by a negligible
reduced Reynolds number, $\epsilon Re$, defined as 
\begin{equation}
\epsilon Re=\epsilon\frac{\tilde{\rho}\tilde{u}^{*}\tilde{h}}{\tilde{\eta}}\ll1.\label{eq:Reduced Re}
\end{equation}
We assume that the Dukhin number, $Du=\tilde{\sigma}_{s}/\tilde{\sigma}_{b}\tilde{l}$, \cite{Dukhin,Lyklema} is small compared to $\epsilon$ 
\begin{equation}
Du=\frac{\tilde{\sigma}_{s}}{\tilde{\sigma}_{b}\tilde{l}}\ll\epsilon,\label{eq:Du}
\end{equation}
where $\tilde{\sigma}_{s}/\tilde{\sigma}_{b}$ is the ratio of surface
to bulk conductivities (see \cite{Lyklema}). As noted by Yariv, \cite{Yariv}
and Khair and Squires, \cite{KhairSquires} this ratio is a length scale (also known
as the healing length) which determines the spatial variations in
electric fields due to non-uniformity in surface conduction. High Dukhin numbers would result in spatial variations in electric field, as well as in variations in concentration which may lead to chemiosmotic flow corrections  \cite{Deryaguin,Prieve,KhairSquires2}. Under our assumption, $ Du\ll\epsilon$, the electric field and the bulk concentration can thus be considered to be uniform throughout the domain.

From order of magnitude analysis of (\ref{eq:Continuity}) and (\ref{eq:Momentum})
we obtain, $\tilde{w}^{\textcolor{blue}{*}}=\epsilon\tilde{u}^{*}$ and $\tilde{p}^{*}=\tilde{\eta}\tilde{u}^{*}/\epsilon^{2}\tilde{l}$.
We note that as the flow is driven by the wall boundary conditions,
the characteristic pressure is independent of the viscosity, $\tilde{p}^{*}=-\tilde{\varepsilon}\tilde{\zeta}^{*}\tilde{E}{}^{*}/\epsilon^{2}\tilde{l}$.
Expanding the velocity field and the pressure in powers of $\epsilon$, \cite{Batchelor} and considering the leading order, $O$($\epsilon^{0}$), the normalized continuity and momentum  equations (\ref{eq:Continuity})
and (\ref{eq:Momentum}) take the form
\begin{subequations} 
\begin{equation}
\frac{\partial u}{\partial x}+\frac{\partial v}{\partial y}+\frac{\partial w}{\partial z}=0,\label{eq:non-continiuty}
\end{equation}
\begin{equation}
\boldsymbol{\nabla}_{\Vert}p=\frac{\partial^{2}\boldsymbol{u_{\Vert}}}{\partial z^{2}}+O(\epsilon Re,\epsilon^{2}),\label{eq:Nond-Momentum}
\end{equation}
\begin{equation}
\frac{\partial p}{\partial z}=O(\epsilon^{3}Re,\epsilon^{2}).\label{eq:Nond-grad p-z}
\end{equation}
\end{subequations}Integrating (\ref{eq:Nond-Momentum}) twice with
respect to $z$ while making use of the boundary conditions (\ref{eq:Boundary-conditions}),
we obtain an expression for the in-plane velocity field
\begin{equation}
\boldsymbol{u}_{\Vert}=\frac{1}{2}z\left(z-1\right)\mathbf{\boldsymbol{\nabla}_{\Vert}}p+z\left(\zeta^{U}-\zeta^{L}\right)\boldsymbol{E_{\Vert}}+\zeta^{L}\boldsymbol{E_{\Vert}},\label{eq:in-plane velocity}
\end{equation}
where $\mathbf{\mathbf{\boldsymbol{\nabla}}_{\Vert}}=\left(\partial/\partial x,\partial/\partial y\right)$
is the two dimensional gradient. Similarly, the perpendicular velocity
can be expressed as 
\begin{equation}
\boldsymbol{u}_{\bot}=z(1-z)\boldsymbol{E_{\Vert}}\cdot\left[z\mathbf{\nabla_{\Vert}}\zeta^{U}+(z-1)\mathbf{\nabla_{\Vert}}\zeta^{L}\right]\boldsymbol{\hat{z}}.\label{eq:perpendicular velocity}
\end{equation}
Defining the mean in-plane velocity, as $\left\langle \boldsymbol{u}_{\Vert}\right\rangle =\intop_{z=0}^{z=1}\boldsymbol{u}_{\Vert}dz,$
and making use of (\ref{eq:Boundary-conditions}), (\ref{eq:non-continiuty})
and (\ref{eq:in-plane velocity}) yields
\begin{equation}
\mathbf{\boldsymbol{\nabla}_{\Vert}\cdot}\left\langle \boldsymbol{u}_{\Vert}\right\rangle =0,\label{eq:div of in-plane mean velocity}
\end{equation}
and
\begin{equation}
\left\langle \boldsymbol{u}_{\Vert}\right\rangle =-\frac{1}{12}\mathbf{\mathbf{\boldsymbol{\nabla}}_{\Vert}}p+\left\langle \zeta\right\rangle \boldsymbol{E_{\Vert}},\label{eq: in-plane mean velocity}
\end{equation}
where $\left\langle \zeta\right\rangle $ is an arithmetic mean value
of the zeta potential on the walls, $\left\langle \zeta\right\rangle =(\zeta^{U}+\zeta^{L})/2.$
Applying the two dimensional divergence to (\ref{eq: in-plane mean velocity}),
and using (\ref{eq:div of in-plane mean velocity}), we obtain an
equation in terms of the pressure only,
\begin{equation}
\mathbf{\mathbf{\nabla_{\Vert}^{\mathrm{2}}}}p=12\boldsymbol{E_{\Vert}}\cdot\mathbf{\boldsymbol{\nabla}_{\Vert}}\left\langle \zeta\right\rangle .\label{eq:Poisson for pressure}
\end{equation}
Similarly, applying the normal component of the curl operator to (\ref{eq: in-plane mean velocity}),
leads to an equation for the stream function,
\begin{equation}
\mathbf{\mathbf{\nabla_{\Vert}^{\mathrm{2}}}}\psi=\left(\boldsymbol{E}_{\boldsymbol{||}}\times\mathbf{\boldsymbol{\nabla}_{\Vert}}\left\langle \zeta\right\rangle \right)\cdot\boldsymbol{\hat{z}},\label{eq:Poisson for the stream function}
\end{equation}
where $\psi(x,y)$ is the averaged stream function related to the velocity field through $\left\langle \boldsymbol{u}_{\Vert}\right\rangle =\left(\partial\psi/\partial y,-\partial\psi/\partial x\right).$

We note that (\ref{eq:Poisson for pressure}) and
(\ref{eq:Poisson for the stream function}) admit an associated gauge
freedom in the choice of zeta potential, which does not affect
the resulting pressure or the flow field. Specifically, for the case
of an electric field acting in the $\boldsymbol{\hat{x}}$ direction,
it follows that adding an arbitrary function $\zeta_{\psi}(y)$, to the zeta potential, $\left\langle \zeta(x,y)\right\rangle \rightarrow\left\langle \zeta(x,y)\right\rangle +\zeta_{\psi}(y)$,
will not modify the resulting pressure in (\ref{eq:Poisson for pressure}).
Similarly, the stream function in (\ref{eq:Poisson for the stream function})
is indifferent to the transformation $\left\langle \zeta(x,y)\right\rangle \rightarrow\left\langle \zeta(x,y)\right\rangle +\zeta_{p}(x)$. 

Eqs. (\ref{eq:Poisson for pressure}) and (\ref{eq:Poisson for the stream function})
are an uncoupled set of Poisson equations for the pressure and the
stream function, and extend the Hele-Shaw equation \cite{HeleShaw}
to include non-uniform EOF. Notably, (\ref{eq:Poisson for pressure})
includes a source term that depends on gradients of zeta potential
which are parallel to the applied electric field, while (\ref{eq:Poisson for the stream function})
relates the vorticity, $\mathbf{\mathbf{\omega=-\nabla_{\Vert}^{\mathrm{2}}}}\psi$,
to changes of zeta potential in a direction normal to the applied
electric field. In particular, in a region where the non-homogeneous
term in (\ref{eq:Poisson for the stream function}) vanishes, the
resulting flow field would be irrotational and a velocity potential
function could be defined. 

In addition to direct calculation of the flow field arising from a certain zeta distribution, we can also use (\ref{eq:Poisson for the stream function}) to obtain
the required zeta potential distribution for achieving a desired
flow field. Without
loss of generality, we may assume that the uniform electric field
is directed along the $\boldsymbol{\hat{x}}$ axis, $\boldsymbol{E_{\Vert}}=E\boldsymbol{\hat{x}}$.
Solving (\ref{eq:Poisson for the stream function})
in terms of a desired stream function we obtain the zeta potential
distribution 
\begin{equation}
\left\langle \zeta(x,y)\right\rangle =\dfrac{1}{E}\mathbf{\mathbf{\int\nabla_{\Vert}^{\mathrm{2}}}}\psi(x,y)dy+\zeta_{p}(x),\label{eq:Inverse zeta}
\end{equation}
where $\zeta_{p}(x)$ is an arbitrary function of the $x$ argument to be determined from the boundary conditions. 

\section{Axially symmetric Zeta Potential Distribution}

Consider the case where the zeta potential distribution acquires a non-zero value, $\zeta_{0}$, in the inner region ($r_{0}<r$)
and vanishes in outer region, $(r_{0}>r)$. In the following, we adopt the superscripts $in$ and $out$ to distinguish between physical
quantities in each one of the two regions. The corresponding zeta potential distribution is given by 
\begin{equation}
	\left\langle \zeta(r)\right\rangle =\zeta_{0}H(r_{0}-r),\label{RadialSymDistribution}
\end{equation}
where $r$ is the distance as measured from the center of our coordinate system which coincides with the center of the disk, $H$ stands for the Heaviside step function, $r_{0}$ is the radius of the disk, which here represents a characteristic length scale $l$, and $\zeta_{0}$ is the constant value of the zeta potential in the disk.

Clearly, the abrupt change in zeta potential value described by (\ref{RadialSymDistribution}) locally violates the lubrication approximation, which assumes negligible gradients in the $x-y$ plane. Nevertheless,
the resulting deviations are expected to be limited to a narrow region, of order $\epsilon$, similar to deviations near the step changes of a cross section in a long and narrow channel \cite{BrothertonDavis}.

For $\boldsymbol{E_{\Vert}}=(E,0)$, (\ref{eq:Poisson for pressure}) takes the following form in polar coordinates
\begin{equation}
	\frac{1}{r}\frac{\partial}{\partial r}\left(r\frac{\partial p}{\partial r}\right)+\frac{1}{r^{2}}\frac{\partial^{2}p}{\partial\theta^{2}}-12E\cos(\theta)\frac{d\left\langle \zeta\right\rangle }{dr}=0,\label{eq:Poisson for pressure on polar}
\end{equation}
where $\theta$ is the azimuthal angle in the plane and $x=r\cos(\theta)$.
The last term in (\ref{eq:Poisson for pressure on polar}) suggests
a solution of the form
\begin{equation}
           p(r,\theta)=\alpha(r)\cos(\theta),
\label{Solution form}
\end{equation}
where $\alpha(r)$ is yet to be determined. Combining (\ref{RadialSymDistribution})-(\ref{Solution form})
yields 
\begin{equation}
	\frac{1}{r}\frac{d}{dr}\left(r\frac{d\alpha(r)}{dr}\right)-\frac{\alpha(r)}{r^{2}}+12E\zeta_{0}\delta(r_{0}-r)=0,
\label{ODE for Alpha}
\end{equation}
where $\delta$ is the Dirac delta function. Solving (\ref{ODE for Alpha}) separately in the inner and outer regions, and requiring regularity yields the following expression for the pressure
\begin{equation}
          p(r,\theta)=\begin{cases}
\left(\dfrac{a^{out}}{r}+b^{out}r\right)\cos(\theta)\hspace{0.1in}\\
b^{in}r\cos(\theta),
\end{cases}
\label{Pressure diploe}
\end{equation}
where $b^{in}$, $b^{out},$ and the dipole strength $a^{out}$ are coefficients to be determined from boundary conditions. Utilizing
(\ref{eq: in-plane mean velocity}), (\ref{RadialSymDistribution}) and (\ref{Pressure diploe}) provides the associated closed-form expression for the flow field
\begin{equation}
	\left\langle \boldsymbol{u}_{\Vert}\right\rangle =\begin{cases}
\dfrac{1}{12}\left[\left(\dfrac{a^{out}}{r^{2}}-b^{out}\right)\cos(\theta)\boldsymbol{\hat{r}}+\left(\dfrac{a^{out}}{r^{2}}+b^{out}\right)\mathrm{sin}(\theta)\boldsymbol{\hat{\theta}}\right]\hspace{0.1in} & r_{0}<r\\
\dfrac{1}{12}\left[\left(-b^{in}+12E\zeta_{0}\right)\cos(\theta)\boldsymbol{\hat{r}}+\left(b^{in}-12E\zeta_{0}\right)\mathrm{sin}(\theta)\boldsymbol{\hat{\theta}}\right]\hspace{0.1in} & r<r_{0}.
\end{cases}
\label{the dipole velocity}
\end{equation}
In the inner region, the pressure changes linearly in the direction of the electric field, which results in a uniform velocity vector
field, explicitly given by
\begin{equation}
	\left\langle \boldsymbol{u}_{\Vert}\right\rangle ^{in}=\left(-\frac{1}{12}\frac{\partial p}{\partial x}+E\zeta_{0}\right)\boldsymbol{\hat{x}}=\left(-\frac{1}{12}b^{in}+E\zeta_{0}\right)\boldsymbol{\hat{x}}.
\label{inner constant velocity}
\end{equation}
The coefficients are determined by demanding continuity of the pressure
and radial velocity component at $r=r_{0}$, leading to 
\begin{equation}
a^{out}=6E\zeta_{0}r_{0}^{2};\quad b^{in}-b^{out}=6E\zeta_{0}.\label{dipole strength+bin+bout}
\end{equation}
The dipole strength $a^{out}$ represents a contribution to the pressure
as a result of the zeta potential in the disk, while the constant
$b^{out}$ represents the magnitude of a uniform flow field in the
$\boldsymbol{\hat{x}}$ direction which stems from an externally applied
pressure gradient in that direction, $b^{out}=\Delta p/\Delta x$
(e.g. for the case of constant pressure at $r\rightarrow\infty$,
$b^{out}=0$). $b^{in}$ is then readily obtained from (\ref{dipole strength+bin+bout}).
The value of $b^{in}-b^{out}$ represents discontinuity in the $\boldsymbol{\hat{\theta}}$
component of the depth-averaged velocity on the surface $r=r_{0}$,
induced by the vorticity sources specified by (\ref{eq:Poisson for the stream function}).
From (\ref{the dipole velocity}) we find that the corresponding jump in the $\boldsymbol{\hat{\theta}}$ component of the velocity (i.e. difference between
its outer and inner values) is $-E\zeta_{0}$. This result can be also
obtained by using the relation $\left\langle u_{\theta}\right\rangle =-\text{\ensuremath{\partial}}\psi/\text{\ensuremath{\partial}}r$
and integrating (\ref{eq:Poisson for the stream function}) along
infinitesimal region around $r_{0}$. 

While it is expected from multipole expansion theorem \cite{Leal} that far
from the non-uniformity the flow field will decay to a dipole \cite{Long}, we note that in this case
the flow exhibits an exact dipole behavior even in immediate vicinity
of the disk.

\begin{figure}[h]
\centering{}\includegraphics[scale=0.9]{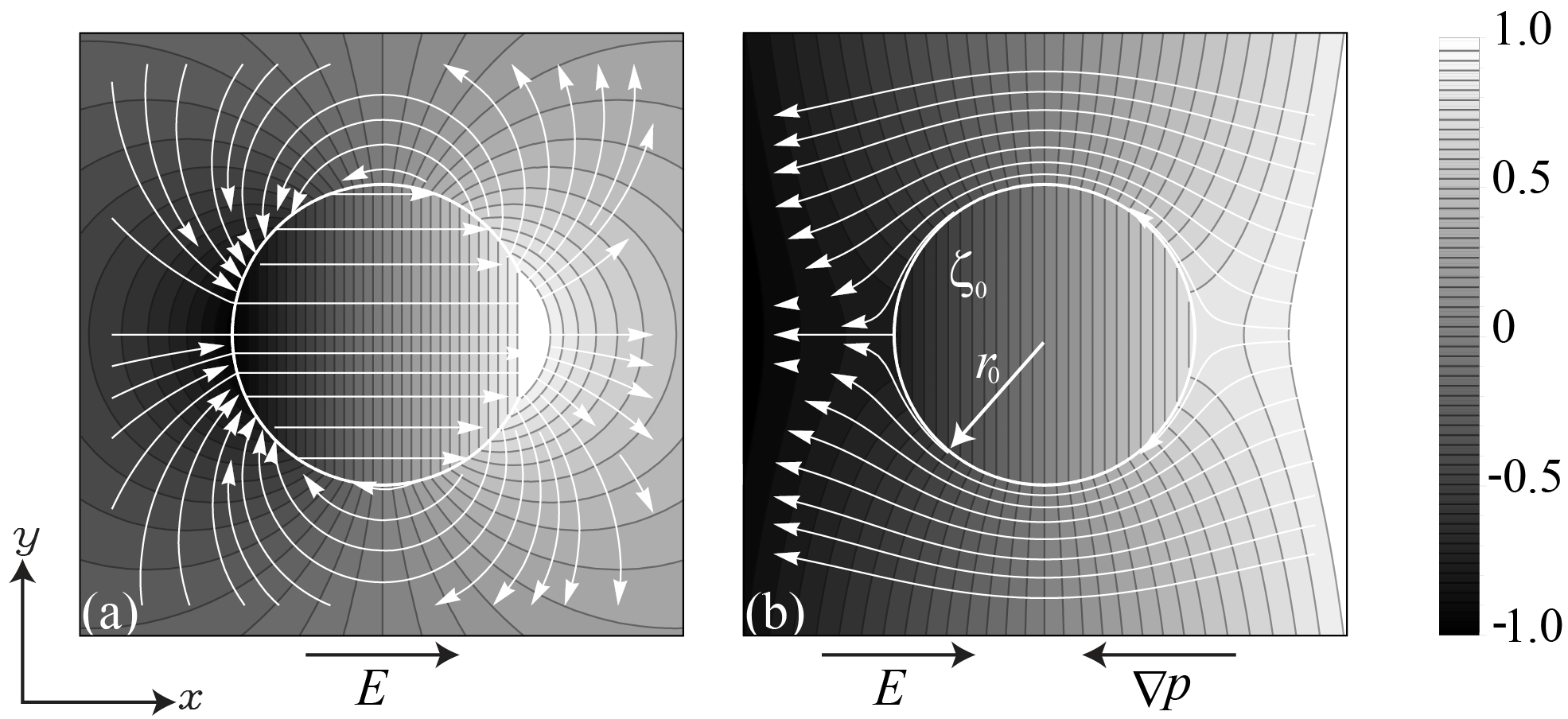}\protect\caption{\textit{The pressure distribution (colormap) and streamlines (white
lines) due to a uniform electric field in the $\boldsymbol{\hat{x}}$ direction, applied to a Hele-Shaw configuration having a non-zero
zeta potential in a disk of radius $r_{0}$. (a) The case of no incident flow. The depth-averaged flow field described by (\ref{the dipole velocity}), results in a uniform flow in the inner region and dipole flow in the
outer region. Note that the points of extremum pressure and vorticity correspond to locations where the source terms in (\ref{eq:Poisson for pressure}) and (\ref{eq:Poisson for the stream function}) attain their minimal
or maximal values. (b) The case where an external pressure gradient (counter-flow) is applied such that the velocity inside the disk vanishes. The result coincides with that of potential flow around a cylinder and is characterized by open streamlines which do not enter the disk region. Both solutions are normalized such that the pressure varies between $\pm 1$. }}
\end{figure}
Figure (2a) presents the streamlines and pressure distribution map of a single electroosmotic dipole for the case of vanishing pressure
far from the disk ($b^{out}=0$). As expected, the boundary conditions in the inner region dictate that the flow is from low pressure to high pressure while the flow in the outer region is from high pressure to low pressure. Interestingly, taking advantage of the uniform flow field in the inner region, we can superpose it with another depth-averaged flow of equal magnitude and opposite direction, to achieve zero net flow in the inner region, i.e. $\left\langle \boldsymbol{u}_{\Vert}\right\rangle ^{in}=0$.
This can be achieved by applying a mean pressure gradient $\Delta p/\Delta x=b^{out}=6E\zeta_{0}$,
or by adding a bias value of $-\zeta_{0}/2$ to the zeta potential everywhere in the domain. Figure (2b) presents the streamlines and
pressure distribution for this case. The corresponding flow field around the disk coincides then with the well-known solution of potential flow past an infinitely long cylinder.

\begin{figure}[h]
\centering{}\includegraphics[scale=0.9]{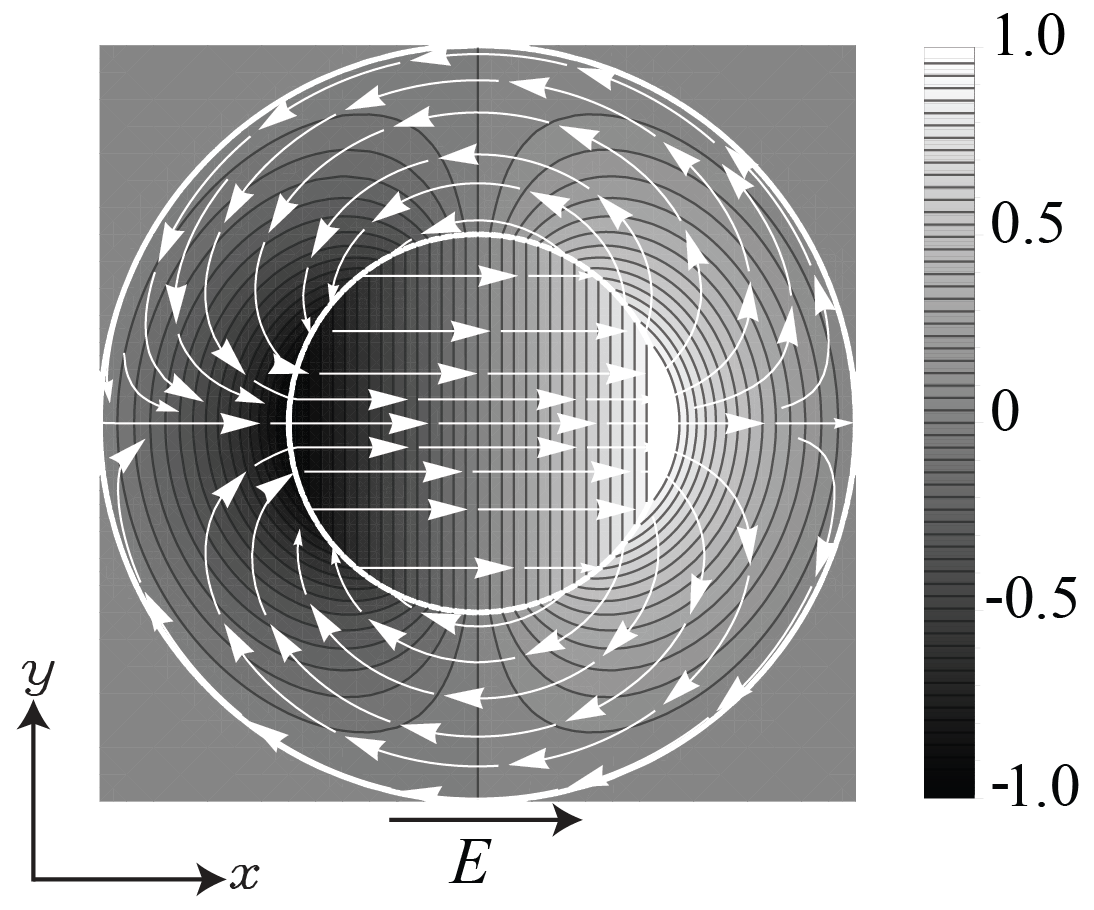}
\protect\caption{\textit{Analytical solution due to superposition of two disks, described by Eqs (\ref{Pressure diploe}) and (\ref{the dipole velocity}), showing the superposition of two disks of radii $r_{0}/2$ and $r_{0}$ with zeta potentials, $\zeta_{0}$ and $-\zeta_{0}/4$, respectively. Colormaps describe the normalized pressure distribution and the white lines show streamlines. The resultant dipole strength vanishes outside the larger disk, giving rise to a pressure and flow which are confined within the outer disk and uniform flow in the inner disk.}}
\end{figure}
Employing the linearity of the governing equations enables to superpose
the basic dipole solution, (\ref{Pressure diploe}), and construct axially symmetric solutions where the zeta potential, $\zeta_{0}(r)$, is a function of the radial coordinate. For the case in which $\zeta_{0}(r)$ vanishes outside a disk of radius $r_{0}$, we obtain the corresponding pressure distribution, $p_{tot}$,
\begin{equation}
           p_{tot}(r,\theta)=\begin{cases}
          \dfrac{12E\cos(\theta)}{r}\int\limits _{0}^{r{}_{0}}\zeta_{0}(r'_{0})r'_{0}dr'_{0} & r_{0}<r\\
          12E\cos(\theta)\left[\dfrac{1}{r}\int\limits _{0}^{r}\zeta_{0}(r'_{0})r'_{0}dr'_{0}+\dfrac{r}{2}\left(\int\limits          
          _{r}^{r_{0}}\zeta_{0}(r'_{0})r'_{0}dr'_{0}\right)\right]\cos(\theta) & r<r_{0},
\end{cases}
\label{Pressure Arbitrary}
\end{equation}
where the solution in the outer region corresponds again to an exact
dipole, having a dipole strength $12E\int\limits _{0}^{r{}_{0}}\zeta_{0}(r'_{0})r'_{0}dr'_{0}$. Figure (3) presents an analytical solution for superposition of two
concentric disks of radii $r_{0}/2$ and $r_{0}$ with zeta potentials
$\zeta_{0}$ and $-\zeta_{0}/4$, respectively. For this special case,
the pressure vanishes outside the larger disk, and results in flow
that is confined to the boundaries of the outer disk, whereas a uniform
velocity field is obtained in the small disk, as expected.

One can also readily apply a combination of disk-shaped regions with constant zeta potential as a tool
to achieve a required flow field in a Hele-Shaw cell with a uniform electric field. Specifically, as an illustration of this approach,
we apply the panel method, commonly used in aerodynamics \cite{KatzPlotkin}, to design EOF around a symmetric NACA 0015 airfoil
profile. For concreteness, we choose the direction of the electric field in our example along the $\boldsymbol{\hat{x}}$
axis, and specify a known pressure gradient, $\Delta p/\Delta x$, along this direction. We place a set of 24 disks, each having a uniform (but potentially different) zeta potential, also along this axis. We then utilize the closed form dipole solution generated by each disk to calculate the velocity vector at 24 points along the airfoil curve. Demanding no-penetration at each of the points, results in a set of linear algebraic equations for the associated $r_{0}$ and $\zeta_{0}$ values of the disks.
\begin{figure}[h]
\centering{}\includegraphics[scale=0.9]{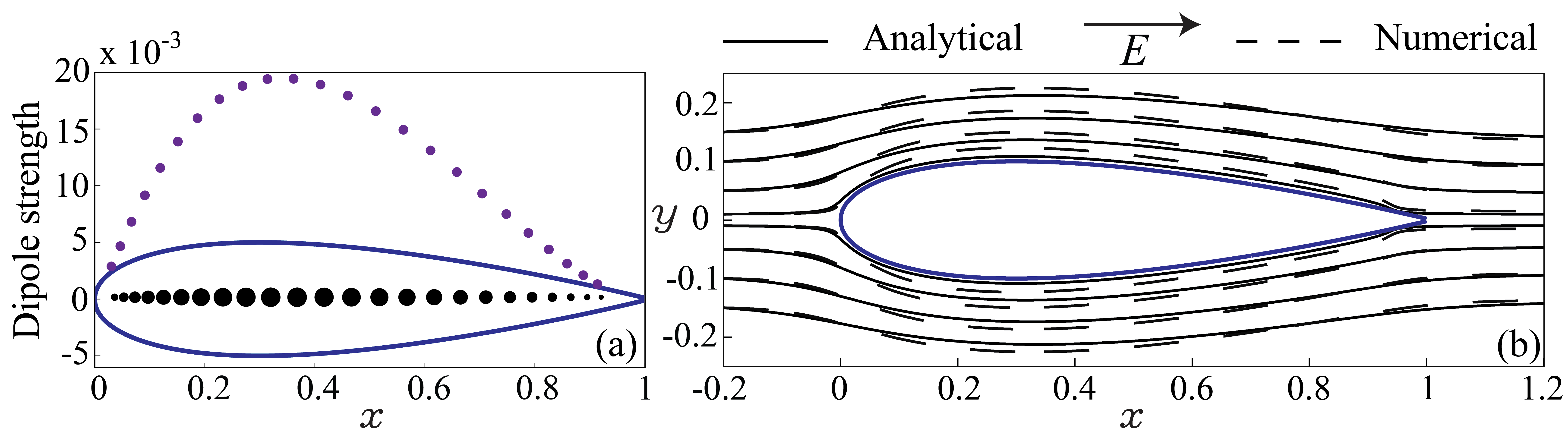}
\protect
\caption{\textit{Superposition of disks with uniform zeta potential enables
creation of complex flow patterns, satisfying no-penetration on desired convex surfaces. (a) 24 circular disks are distributed along the }\textit{{$\boldsymbol{\hat{x}}$axis}}
\textit{. The non-dimensional dipole strength of each of the disks is set such that a no-penetration boundary condition is obtained
over a NACA 0015 airfoil. (b) Comparison of analytical (solid lines) and numerical results (dashed lines) of the flow field obtained from
the distribution. Analytical results were obtained from 2D Hele-Shaw model, whereas the numerical results are depth-averaged from a direct three dimensional numerical solution using Comsol Multiphysics 4.4. }}
\end{figure}

Figure (4a) shows the intensity of each of the disks as well as their location along the $\boldsymbol{\hat{x}}$
axis. Figure (4b) presents the streamlines (solid lines) obtained
from the analytical solution. To validate the results, we performed a three dimensional direct numerical simulation using Comsol Multiphysics 4.4, in which the Stokes flow equations are solved, coupled to Helmholtz– Smoluchowski boundary conditions. Our grid consisted of 120,504 $2^{nd}$ order unstructured prism elements, and all solutions converged by at least seven orders of magnitude from a zero velocity initial condition. We used equally sized disks having zeta potential values as determined by the analytical solution.
The velocity field is depth-averaged, and streamlines of this numerical solution are also presented in the figure (dashed lines), showing good agreement with the analytical result. 

\section{Uniform zeta-potential along a Thin Finite Line}

Eqs. (\ref{eq:Poisson for pressure}) and (\ref{eq:Poisson for the stream function}) are both Poisson-type equations for the pressure and vorticity, in which gradients in zeta potential serve as source terms. The respective
dot-product and curl operators in these source terms suggest that the solutions would depend on the geometry of surface
patterning. To further highlight the roles of these source terms, let us consider two simple cases of EOF arising from a thin finite
line with uniform zeta potential which is either aligned with the electric field or perpendicular to it. In both cases, we model
the lines using the linearity of the governing equations, and sum up the pressure contributions of dipoles having a strength per unit length $a^{out}$
\begin{equation}
	dp(x,y)=\dfrac{a^{out}(x-x_{0}(l))}{(x-x_{0}(l))^{2}+(y-y_{0}(l))^{2}}dl,\label{PressureElement}
\end{equation}
where the dipole position is parametrized by $l$, and $dl$ is an
infinitesimal interval along the curve $l$. We then use (\ref{eq: in-plane mean velocity})
to derive the fluid velocity profile outside the dipole distribution. 

For the case of a line of dipoles directed along the electric field,
the pressure distribution and the mean velocity are given by
\begin{subequations}
\begin{equation}
           p(x,y)=a^{out}\ln\left(\dfrac{r_{-}}{r_{+}}\right),\label{Pressure Two Monopoles}
\end{equation}
\begin{equation}
	\left\langle \boldsymbol{u}_{\Vert}\right\rangle =-\frac{a^{out}}{12}\left(\dfrac{\boldsymbol{\hat{r}_{-}}}{r_{-}}-\dfrac{\boldsymbol{\hat{r}}_{+}}{r_{+}}\right),\label{Velocity Two Monopoles}
\end{equation}
\end{subequations}
where $\boldsymbol{\hat{r}_{\pm}}$
are unit vectors pointing from the observation point $(x,y)$ to the
edge points $(x_{-},y_{0})$ and $(x_{+},y_{0})$ , and $r_{\pm}=\sqrt{\left(x-x_{\pm}\right)^{2}+\left(y-y_{0}\right)^{2}}$
are the corresponding distances between these points. For a case where
the line connects the points $(x_{0},y_{-})$ and $(x_{0},y_{+})$
, and is perpendicular to electric field, the pressure distribution
and the mean velocity are given by
\begin{subequations}
\begin{equation}
           p(x,y)=a^{out}\left(\tan^{-1}\left(\dfrac{y-y_{-}}{x-x_{0}}\right)-\tan^{-1}\left(\dfrac{y-y_{+}}{x-x_{0}}\right)\right),\label{Pressure Two Vortices}
\end{equation}
\begin{equation}
	\left\langle \boldsymbol{u}_{\Vert}\right\rangle =\frac{a^{out}}{12}\left(\dfrac{\boldsymbol{\hat{\theta}}_{+}}{r_{+}}-\dfrac{\boldsymbol{\hat{\theta}}_{-}}{r_{-}}\right).\label{Velocity TwoVortices}
\end{equation}
\end{subequations}where $\boldsymbol{\hat{\theta}}_{\pm}$ are unit
vectors which perpendicular to $\boldsymbol{\hat{r}_{\pm}}$.

The solutions (\ref{Pressure Two Monopoles}) and (\ref{Velocity Two Monopoles})
can be interpreted as a sink and a source, located at the edges of
the uniform zeta potential line. Figure (5a) presents the resulting
pressure and velocity field for this case. Clearly, the apparent violation
of continuity, $\mathbf{\boldsymbol{\nabla}_{\Vert}\cdot}\left\langle \boldsymbol{u}_{\Vert}\right\rangle =0$, at the two points is an artifact of the fact that (\ref{Pressure Two Monopoles})
and (\ref{Velocity Two Monopoles}) represent the flow outside the array of dipoles, and additional flow that connects $(x_{-},y_{0})$ and $(x_{+},y_{0})$ exists within this array. Note that in this
case all zeta potential gradients are in the direction of the electric field and thus, consistent with the streamlines equation
(\ref{eq:Poisson for the stream function}), the flow field (\ref{Velocity Two Monopoles})
is free of vorticity. The solutions (\ref{Pressure Two Vortices}) and (\ref{Velocity TwoVortices}) are effectively equivalent to a pair
of equal strength and opposite sign vortices, centered at the edges of the dipole line. Figure (5b) presents the exact analytical solution for the pressure and velocity, as given by (\ref{Pressure Two Vortices}) and (\ref{Velocity TwoVortices}). Consistent with (\ref{eq:Poisson for pressure}), high pressure gradients are formed between the two sides of the line, where the zeta potential
changes in a direction parallel to electric field. At the line's edges the pressure difference results in circulatory flow, again
consistent with (\ref{eq:Poisson for the stream function}) which predicts vorticity where the zeta potential changes in a direction
normal to electric field (in this case, the uniform zeta line terminating into the zeta-free surface). The circulation, defined as the line
integral, $\Gamma_{\pm}=\oint\left\langle \boldsymbol{u}_{\Vert}\right\rangle \cdot d\boldsymbol{l}$,
along a closed path around each edge point coincides (up to a factor of $\pi/6$ ) with individual dipole strength $\Gamma_{\pm}=\pm\pi a^{out}/6$. Clearly, $\Gamma_{+}+\Gamma_{-}=0$, i.e., the total circulation remains
zero. More generally, from (\ref{eq:Poisson for the stream function}) it follows that the total vorticity generated by non-homogeneity of an arbitrary shape which hosts a constant zeta potential must vanish: the component of the zeta potential gradient perpendicular
to the electric field is $\left\langle \zeta\right\rangle \mathrm{sin}(\theta)$, and thus its line integral along the closed curve must vanish.
\begin{figure}[h]
\centering{}\includegraphics[scale=0.9]{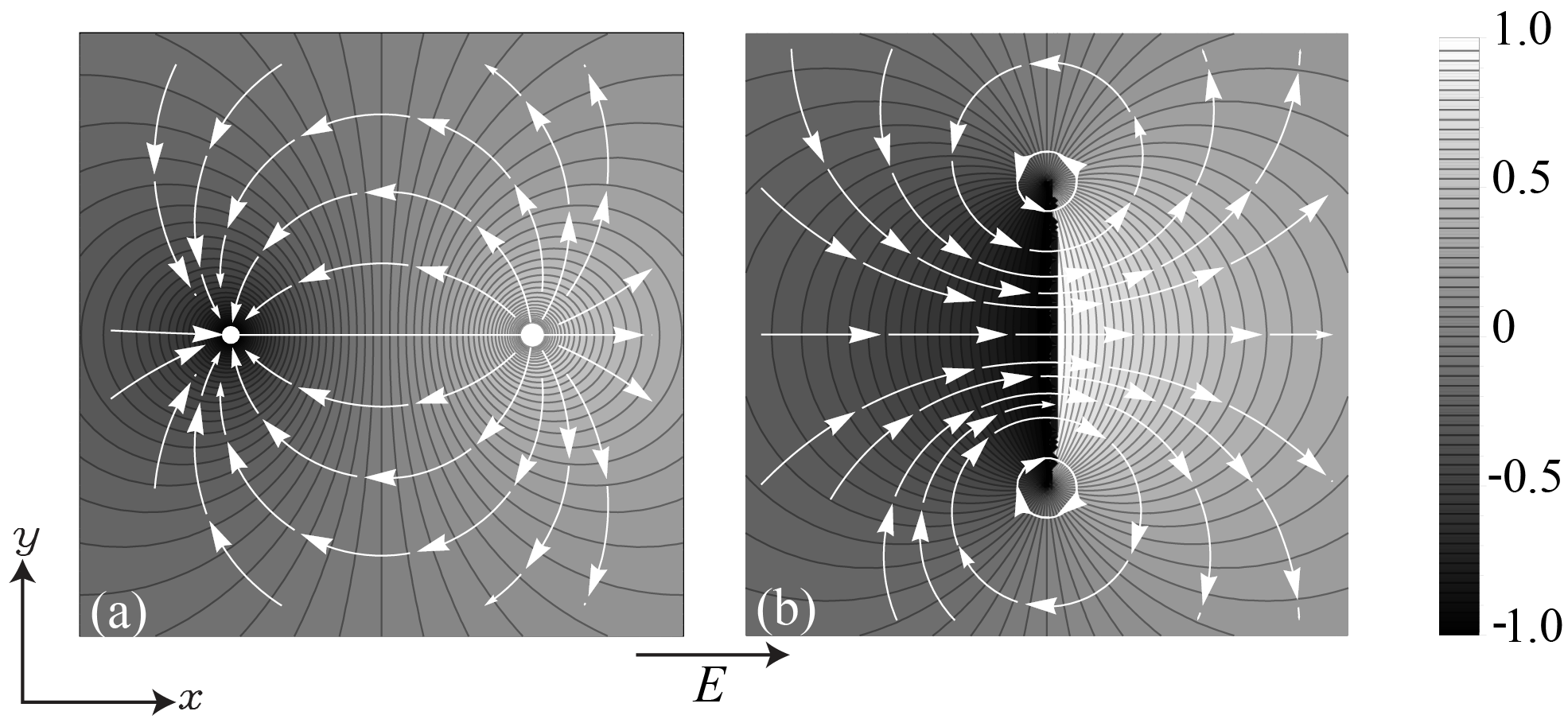}\protect\caption{\textit{Analytical solution showing the flow field due to a line of
uniform zeta potential aligned (a) along the direction of the
electric field and (b) perpendicular to the electric field. Each solution
is obtained from superposition of dipoles, according to (\ref{PressureElement}).
Colormaps describe the normalized pressure distribution and the white
lines show streamlines. In configuration (a) gradients in zeta potential
are all in the direction of the electric field, resulting in source/sink
terms at the edges of the line for the pressure equation (\ref{eq:Poisson for pressure}).
The streamline equation (\ref{eq:Poisson for the stream function}),
on the other hand, remains homogenous and no vorticity is formed.
In (b), the line’s edges form a zeta potential gradient perpendicular
to the electric field, resulting in vorticity and localized circulation
around each for the edges. In addition, a pressure difference forms
across the line, reflecting the fact the zeta potential changes
in a direction parallel to the direction of the electric field.}}
\end{figure}

\section{Zeta Potential Distribution Defined by a Required Flow Field and its Application to Create Complex Microfluidic Networks}

In sections IV and V the zeta potential distribution geometry was predefined and then solved to obtain the flow field. More complex flow fields were obtained by superposing solutions. Here we focus on obtaining the zeta potential distribution required in order to establish a desired flow field. This approach enables to generate
depth-averaged flow fields which cannot be created using external pressure actuation at the boundaries. This is done by setting a desired stream function, and using (\ref{eq:Inverse zeta}) to calculate the required zeta potential distribution. Without loss of generality,
we here assume that the uniform electric field is directed along the $\boldsymbol{\hat{x}}$ axis, $\boldsymbol{E_{\Vert}}=E\boldsymbol{\hat{x}}$.

We illustrate this approach by considering flow in a square domain $[0,1]\times[0,1]$, where $x=0,\,1$ and $y=0$ correspond to impermeable walls, and $y=1$ corresponds to a surface of constant pressure (see
figure (6a)). An example of such a flow is given by the stream function, 
\begin{equation}
\psi(x,y)=y\mathrm{sin}(\pi x),\label{eq:Stream func-1}
\end{equation}
describing a U-shaped flow (see solid streamlines in figure (6a))
in which the flow enters and exits the same $y=1$ face. Utilizing
(\ref{eq:Inverse zeta}) and (\ref{eq:Stream func-1}) yields the
corresponding expression for the zeta potential
\begin{equation}
\left\langle \zeta(x,y)\right\rangle =-\dfrac{\pi^{2}}{2E}y\mathrm{^{2}sin}(\pi x)+\zeta_{p}(x).\label{eq:zeta-1}
\end{equation}
Defining the normal and tangential unit vectors to the domain boundaries
as $\boldsymbol{\hat{n}}$ and $\boldsymbol{\hat{t}}$, respectively,
the no-penetration boundary condition, $\mathbf{\boldsymbol{\nabla}_{\Vert}}\psi\cdot \boldsymbol{\hat{t}}=\left\langle \boldsymbol{u}_{\Vert}\right\rangle \cdot\boldsymbol{\hat{n}}=0$
is automatically satisfied by the stream function (\ref{eq:Stream func-1})
on $x=0,\,1$ and $y=0$. To satisfy the constant pressure condition,
$\mathbf{\boldsymbol{\nabla}_{\Vert}}p\cdot \boldsymbol{\hat{t}}=0,$ on the face
$y=$1, we utilize the gauge freedom in the choice of the function
$\zeta_{p}(x)$. Substituting this condition into (\ref{eq: in-plane mean velocity})
results in $\left\langle \boldsymbol{u}_{\Vert}\right\rangle \cdot\boldsymbol{\hat{x}}=\left\langle \zeta\right\rangle \mid_{y=1}E$.
Expressing the velocity in terms of the stream function, and substituting
the expression for zeta potential (\ref{eq:zeta-1}) leads to
\begin{equation}
	\zeta_{p}(x)=\frac{1}{E}\left(1+\frac{\pi^{2}}{2}\right)\mathrm{sin}(\pi x).\label{eq: zeta-0}
\end{equation}
Consequently, the corresponding zeta potential distribution and
the pressure are \begin{subequations}
\begin{equation}
	\left\langle \zeta(x,y)\right\rangle =\dfrac{1}{2E}\left[\pi^{2}\left(1-y^{2}\right)\mathrm{sin}(\pi x)+2\right]\mathrm{sin}(\pi x),\label{eq:zeta final}
\end{equation}
\begin{equation}
             p(x,y)=6\pi\left(y^{2}-1\right)\mathrm{cos}(\pi x),\label{eq: pressure returning channel}
\end{equation}
\end{subequations}and are presented in figures (6b) and (6c), respectively.

To validate the results, we compare the desired flow field (\ref{eq:Stream func-1})
with the flow field obtained by numerically solving a three dimensional
case in which the resulting zeta potential (\ref{eq:zeta final}) is used in the slip boundary conditions on both the lower and upper
walls. To solve the corresponding Stokes equations, we used the creeping
flow package in Comsol Multiphysics 4.4. Streamlines of the depth-averaged
velocity field are presented in figure (6a) (dashed lines), showing good agreement with the desired flow. 
\begin{figure}[h]
\centering{}\includegraphics[scale=0.45]{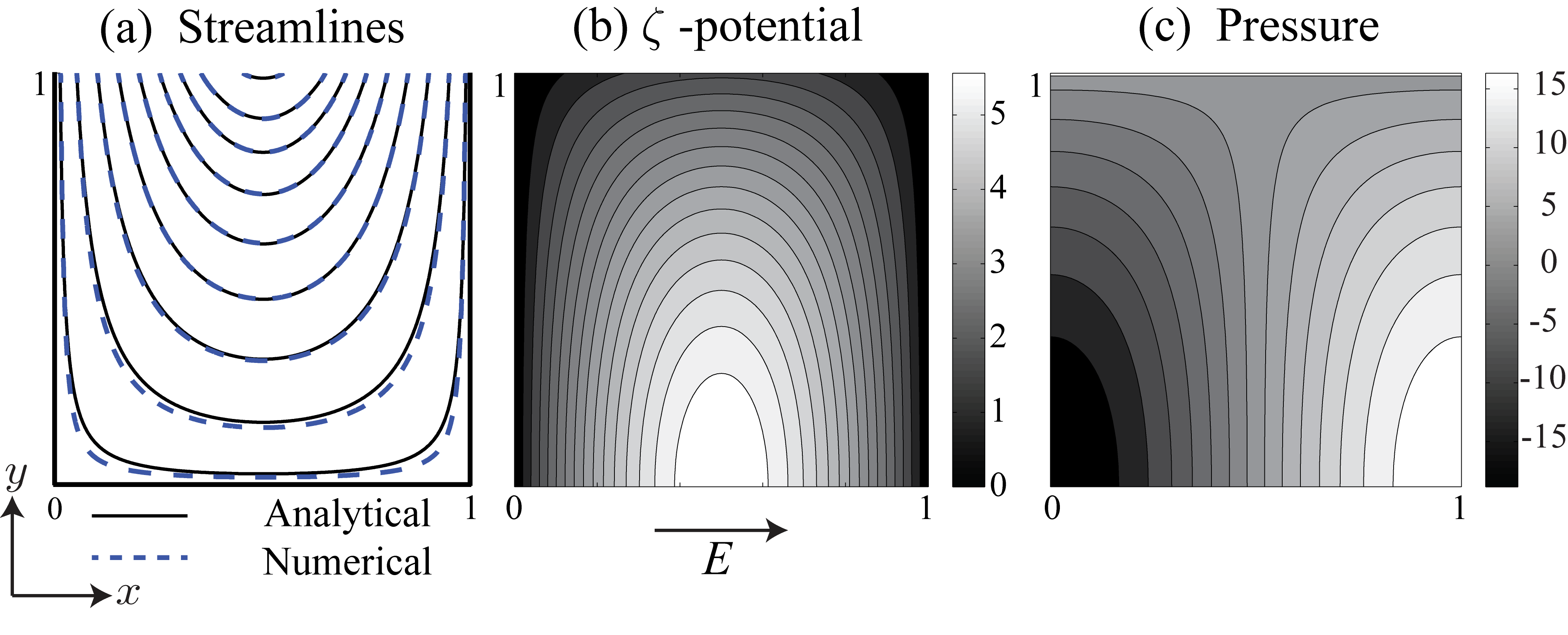}\protect\caption{\textit{Demonstration and numerical validation of the inverse problem,
in which the zeta potential distribution is calculated from a
desired flow field. (a) Comparison of the desired streamlines (solid
lines) with the depth-averaged streamlines obtained from a three dimensional
numerical simulation (dashed lines). (b) The analytical zeta potential
distribution, (\ref{eq:zeta final}), obtained in order to generate
the desired flow field. (c) The resulting pressure distribution (\ref{eq: pressure returning channel}).
All calculations are performed using $E=1$. }}
\end{figure}

We now turn to illustrate the use of (\ref{eq:Inverse zeta}) to engineer
complex flow fields of the type encountered in microfluidic devices.
These flows are typically confined to narrow channels composed of turns, junctions and straight segments. Importantly,
in our approach there are no real walls that confine the liquid, and yet using appropriate zeta potential, the fluid achieves desired
directionality within specific regions of interest and its velocity
strongly decays outside them. We show that under certain conditions the corresponding zeta potential distribution necessary to generate complex flow fields of such type admits modularity, i.e. can be represented as a sum of simple and basic distributions which allow sufficiently smooth matching.

First we wish to determine the zeta potential necessary to generate a flow field along a curve $y=f(x)$, under a homogeneous electric field. We set a stream function of the form 
\begin{equation}
\psi(x,y)=F(y-f(x))\equiv F(\chi),\label{eq: F(y-f(x))}
\end{equation}which ensures that $\psi$ has a constant value along $y=f(x)$ for any
choice of $F$. The variable $\chi=y-f(x)$ parametrizes different
streamlines according to their displacement along the $\boldsymbol{\hat{y}}$
axis. By definition of the stream function, and using (\ref{eq: F(y-f(x))}),
the velocity components can be expressed as 
\begin{equation}
	\left\langle u\right\rangle =\frac{\partial\psi}{\partial y}=F'(\chi),\quad\left\langle v\right\rangle =-\frac{\partial\psi}{\partial x}=f'(x)F'(\chi).\label{eq: F(y-f(x))-1}
\end{equation}
To determine the zeta potential necessary to generate the flow field (\ref{eq: F(y-f(x))-1}), we substitute the stream function (\ref{eq: F(y-f(x))}) into the relation (\ref{eq:Inverse zeta}) which yields the following closed form expression for the necessary zeta potential
\begin{equation}
	\left\langle \zeta(x,y)\right\rangle =\dfrac{1}{E}\left(F'(\chi)+f'(x)^{2}F'(\chi)-F(\chi)f''(x)\right)+\zeta_{p}(x),\label{eq:zeta F}
\end{equation}
where $\zeta_{p}(x)$ is an arbitrary function. Hereafter we focus
on a case where $\zeta_{p}(x)=0.$ Utilizing (\ref{eq: F(y-f(x))-1})
and polar representation of the flow field components, $\left\langle u\right\rangle =\left\langle \boldsymbol{u}_{\Vert}\right\rangle \cos(\theta)$
and $\left\langle v\right\rangle =\left\langle \boldsymbol{u}_{\Vert}\right\rangle \mathrm{sin}(\theta)$,
(\ref{eq:zeta F}) is rewritten as 
\begin{equation}
\left\langle \zeta(x,y)\right\rangle =\dfrac{1}{E\cos^{2}(\theta)}\left(\left\langle u\right\rangle -F(\chi)\frac{d\theta}{dx}\right).\label{eq:zeta F-rewritten}
\end{equation}
This expression allows to directly obtain the zeta potential required in order to produce the desired flow, described by the stream function (\ref{eq: F(y-f(x))}). In regions where the flow direction changes significantly the second term in (\ref{eq:zeta F-rewritten}) dominates over the first term, and vice versa in regions where the flow field is nearly unidirectional. In particular, in regions with uniform flow the zeta potential admits a simple expression given by $\left\langle u\right\rangle /E\cos^{2}(\theta)=\left\langle \boldsymbol{u}_{\Vert}\right\rangle /E\cos(\theta)$. Also, (\ref{eq:zeta F-rewritten}) shows that the zeta potential
required to generate flow in a direction perpendicular to electric
field (i.e. $\left\langle u\right\rangle =0$), diverges and is thus non-physical. 

A convenient function to define a flow field in which the velocity is confined to a narrow region around the streamline $y=f(x)$ is 
\begin{equation}
	\psi(x,y;\theta_{in},\theta_{out})=\mathrm{tanh}\left(\gamma\left(y-f(x;\theta_{in},\theta_{out}\right)\right).\label{eq:Stream-tanh}
\end{equation}
This function exhibits significant gradients in a region of width
$1/\gamma$ around the streamline. Outside of this region it quickly
reaches a negligible gradient, and thus negligible flow.

For creation of stream function modules, which are useful for the
construction of microchannel segments, it is convenient to choose
the function $f$ as 
\begin{equation}
          f(x;\theta_{in},\theta_{out})=\mathrm{tan\left(\theta_{\mathit{in}}\right)}\frac{x}{\left(1+\mathrm{exp}\left(\beta x\right)\right)^{n}}+\mathrm{tan\left(\theta_{\mathit{out}}\right)}\frac{x}{\left(1+\mathrm{exp}\left(-\beta x\right)\right)^{n}}
\label{eq:Stream-tanh-1}
\end{equation}
This function describes a smooth approximation of a bi-linear curve,
which approaches $y=\mathrm{tan\left(\theta_{\mathit{out}}\right)}x$
as $x$ tends to $+\infty$ and to $y=\mathrm{tan\left(\theta_{\mathit{in}}\right)}x$
as $x$ tends to $-\infty$ . $n$ and $\beta$ are positive parameters
which set the exact shape of the curve near the bending point $(0,0)$.

\begin{figure}[h]
\centering{}\includegraphics[scale=0.5]{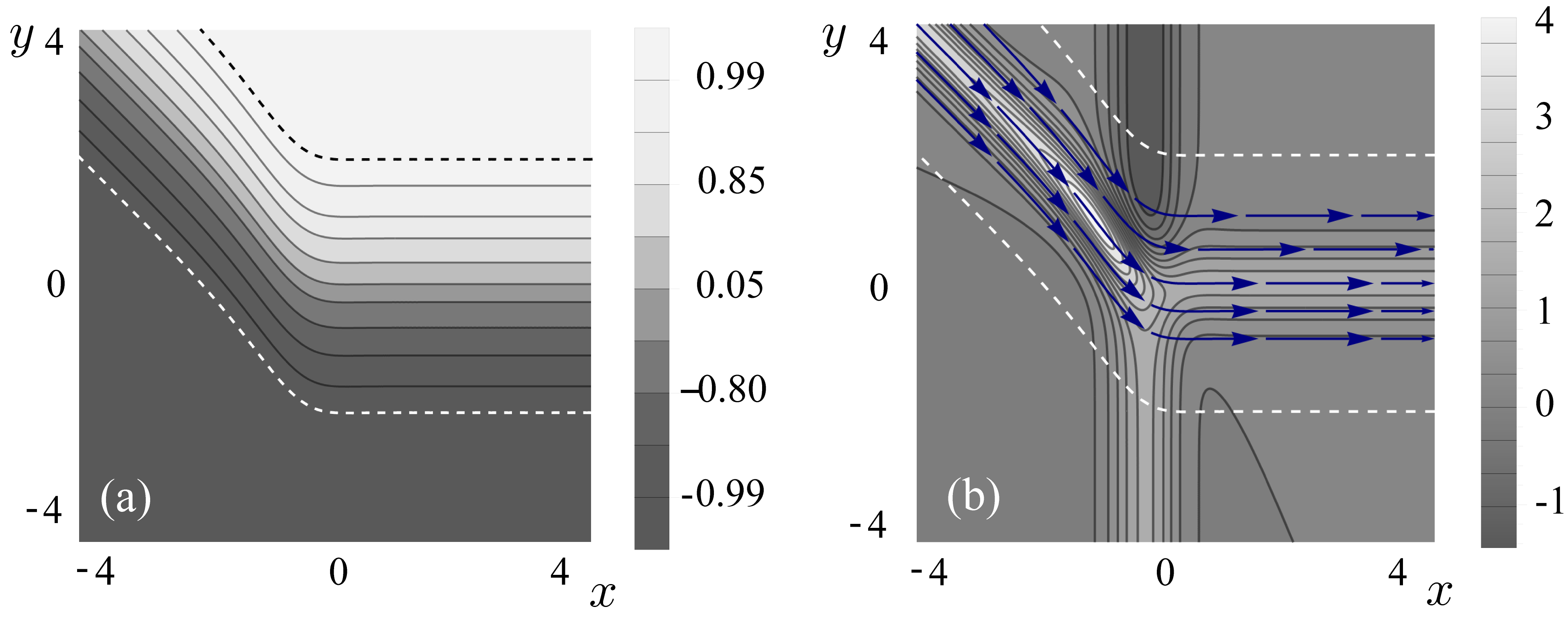}\protect\caption{\textit{Analytical example of zeta potential calculation for a bent channel segment.
(a) Colormap of the stream function $\psi(x,y;-\pi/4,0)$ given by
(\ref{eq:Stream-tanh}), which describes incoming flow at an angle
$-\pi/4$ and horizontal outgoing flow. Here $\beta=\gamma=2,\,n=4,\,E=1$.
The stream function is constructed of two regions having negligible
gradients (and thus negligible flow), which are connected through
a steep gradient region of width $1/\gamma$. The dashed lines correspond
to streamlines on which the velocity drops to $1\%$ of its maximum
value, thus defining the outer boundaries of a channel. (b) Colormap
of the zeta potential, necessary to generate the desired flow field
(blue streamlines), obtained by inserting the corresponding stream
function into (\ref{eq:zeta F}).The pattern shows two regions of
opposite values of zeta potential which extend along the line $x=0$,
towards positive and negative directions of the $\boldsymbol{\hat{y}}$
axis. This contribution stems from the term $-F(\chi)f''(x)$ in
(\ref{eq:zeta F}), and occurs in the region where $f(x)$ admits
its highest second derivative. Two other regions which host significant
values of zeta potential correspond to zones with nearly unidirectional
flow, represented by the first term in (\ref{eq:zeta F-rewritten}).}}
\end{figure}

Figure (7a) presents the contour map of the stream function, $\psi(x,y;-\pi/4,0)$.
The streamlines describe an incoming fluid flowing at an angle $-\pi/4$,
a turn region around the origin and outgoing flow along the $\boldsymbol{\hat{x}}$
axis (zero angle). The region which supports a significant flow velocity
is enclosed between the dashed lines, and its width is of order $1/\gamma$.
The magnitude of the flow velocity on the dashed lines drops to $1\%$
of its peak value. Figure (7b) shows several streamlines (blue arrows),
together with the zeta potential described by (\ref{eq:zeta F-rewritten}),
necessary to generate this flow field. The second derivative of the
function $\psi(x,y;-\pi/4,0)$ acquires significant values around
the line $x=0$, which results in a significant contribution of the
$-F(\chi)f''(x)=-\psi(x,y)f''(x)$ term in (\ref{eq:zeta F}). Since
$f''(x)$ is independent of $y$, and since the stream function values
are constant far from the channel region, this results in zeta potential
values of opposite signs at $x=0$, which extend uniformly toward
$y=\pm\infty$. Far from the line $x=0$, where the flow has a nearly
uniform directionality, the first term in (\ref{eq:zeta F-rewritten})
dominates and the resulting zeta potential distribution is uniform
along the incoming and outgoing flow directions. 

More complex flows can be created by joining of multiple stream function
modules. Figure (8a) shows the stream function, $1/2\left(\psi(x,y;-\pi/4,0)-\psi(x,y;+\pi/4,0)\right)$,
obtained by superposition of two stream functions described by (\ref{eq:Stream-tanh}).
In such superpositions, the resulting flux is thus the sum of individual
fluxes. This allows, for example, constructing junctions where multiple
incoming streams are combined into a single one. In the example shown
in figure (8a), the resulting flow consists of incoming streamlines
along the angles $-\pi/4$ and $\pi/4$ towards the origin, and a
combined outgoing flow along the $\boldsymbol{\hat{x}}$ axis. We
note that the total flux in the channels can be controlled by modifying
the value of $\psi^{(up)}$-$\psi^{(down)}$, whereas the
relative flux of each of the streams can be dictated by modifying
the value $\psi^{(0)}$ for fixed values of $\psi^{(up)}$ and $\psi^{(down)}$.
Figure (8c) shows the zeta potential necessary to generate this
flow field, obtained by substituting the corresponding stream function
into (\ref{eq:zeta F-rewritten}). Note that following the same steps
one can consider a linear combination of $N$ basic functions
$\psi(x,y;\theta_{n},0)$  which generates a
converging flow from $\theta_{n}$ incoming directions and a combined flow
along a single outgoing direction. Figure (8b) presents streamlines obtained from matching the
stream functions $\psi(x-x_{1},y;-\pi/4,0)$ and $\psi(x-x_{2},y;+\pi/4,0)$,
(with $x_{2}=-x_{1}=3.5$), along the common boundary $x=0$. Matching
allows the construction of extended channels in which the flux in
one module
\clearpage
\begin{figure}[t!]
\centering{}\includegraphics[scale=0.38]{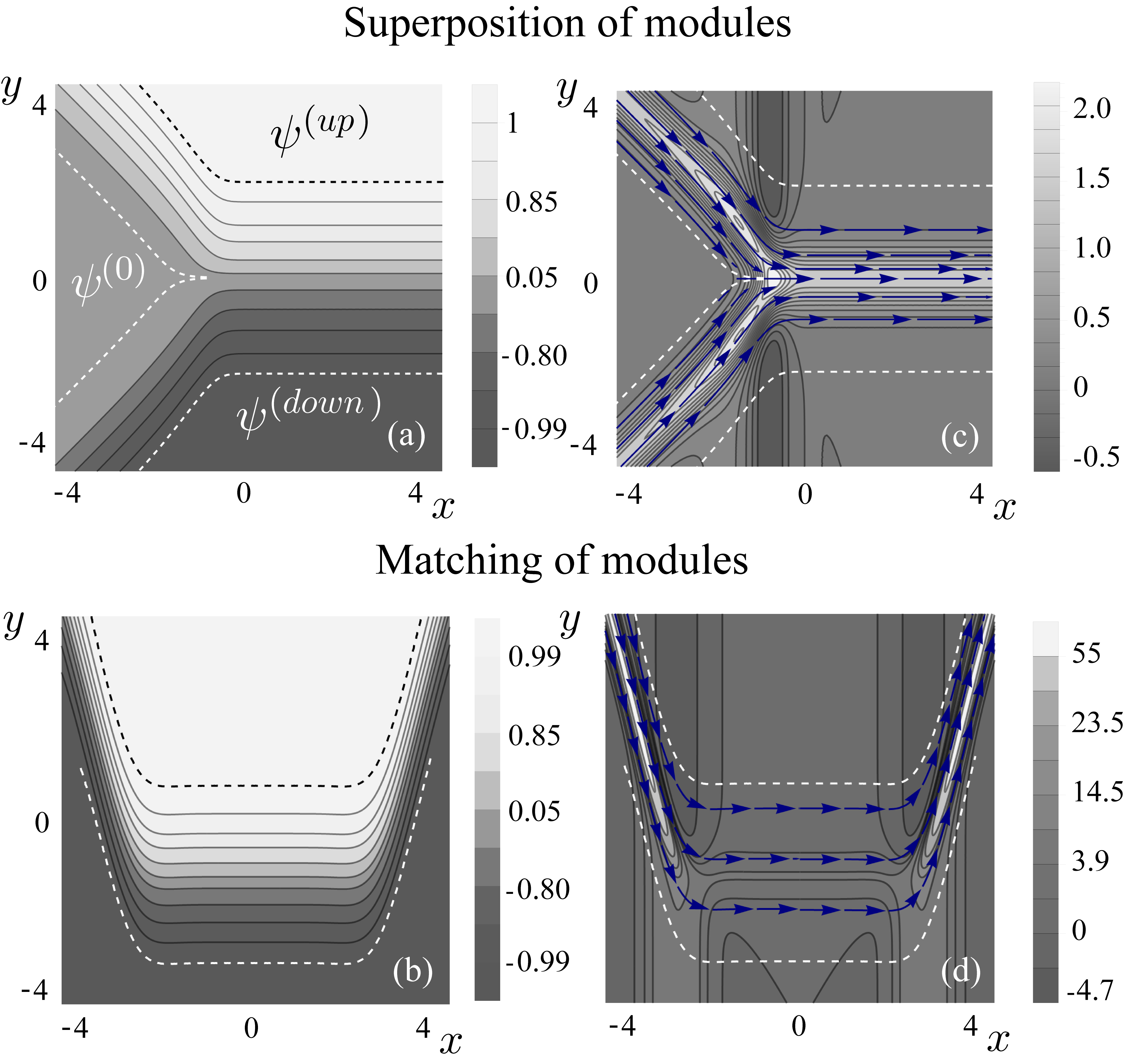}
\protect\caption{\textit{Analytical examples showing superposition and matching of
two basic modules. (a) Superposition assembly of the functions $1/2\psi(x,y;-\pi/4,0)$ and $-1/2\psi(x,y;+\pi/4,0)$
forming a Y-junction in which incoming flows are directed at an angle $\pm\pi/4$, and combine into a single outgoing flow along the $\boldsymbol{\hat{x}}$ axis. The value of $\psi^{(up)}$-$\psi^{(down)}$ determines
the total flow rate in the channel, whereas modifying the value of $\psi^{(0)}$ allows to control the fraction carried by each of the
two incoming flows. (b) Stream function of two modules: $\psi(x-x_{1},y;-\pi/4,0)$
for $x<0$ and $\psi(x-x_{2},y;+\pi/4,0)$ for $x>0$ (with $x_{2}=-x_{1}=3.5$),
matched along the line $x=0$, forming U-shaped streamlines. (c,d) Colormaps of the zeta potential distributions necessary to generate the desired flow field (blue arrows), obtained by substituting the stream functions corresponding to (a,b) into (\ref{eq:zeta F}). In all subfigures the magnitude of flow velocity on the dashed lines drops to $1\%$ of its’ peak value and the parameters which enter (\ref{eq:zeta F}), (\ref{eq:Stream-tanh}) and (\ref{eq:Stream-tanh-1}) are given by $\beta=\gamma=2,\,n=4,\,E=1.$}}
\end{figure}
\hspace{-0.25in} is preserved in the next. In this assembly, the resulting flow consists of an incoming stream along a straight line with slope
$-\mathrm{tan}\left(2\pi/5\right)$, a short horizontal flow parallel
to the $\boldsymbol{\hat{x}}$ axis, and outgoing flow along a straight
line with a slope $\mathrm{tan}\left(2\pi/5\right)$. To join two
such elements, the values of $\zeta$ and its derivative along the
common boundary should ideally be matched.  (\ref{eq:zeta F-rewritten})
shows that the zeta potential of streamlines directed along the
electric field (zero slope), depends solely on the velocity value and
angle and thus can be used for outgoing flow from one domain and incoming flow to another.
\begin{figure}[b]
\centering{}\includegraphics[scale=0.8]{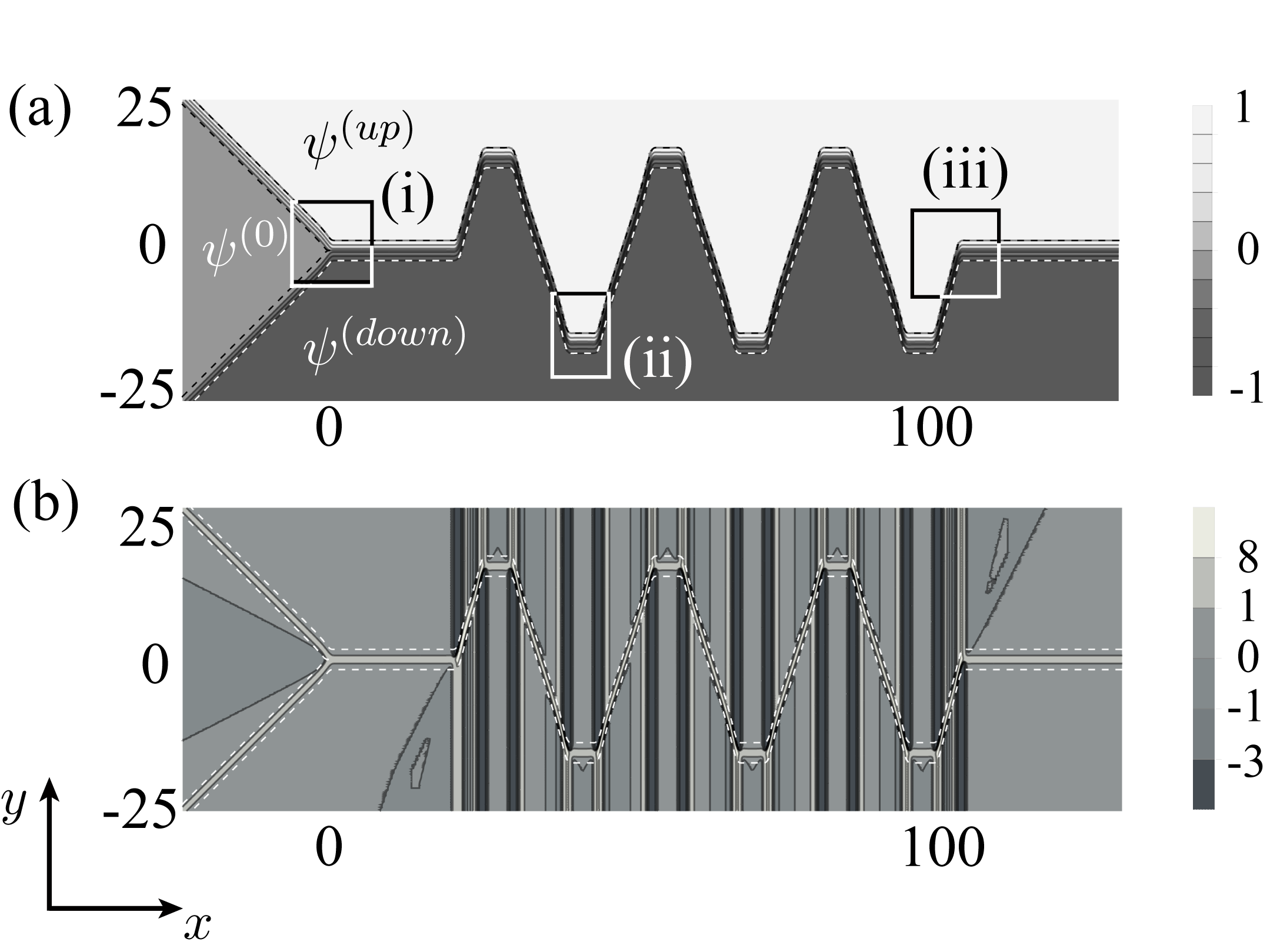}
\protect\caption{\textit{Analytical example demonstrating the use of matching and superposition
of basic modules for creation of a microfluidic network. (a) and (b) show the desired streamlines pattern and the zeta potential distribution necessary to generate the corresponding flow field, respectively. The microfluidic network is composed of a Y-junction, created by superposition assembly of  $1/2\psi(x,y;+\pi/4,0)$ and $-1/2\psi(x,y;-\pi/4,0)$, followed by a matching assembly of $\psi(x-x_{i},y-f_{i};-\pi/4,0)$ and $\psi(x-x_{i},y-f_{i};+\pi/4,0)$ where the index $i$ enumerates
the bending points $\left(x_{i},f_{i}\right)$ of different basic modules, and their matching boundary lines, $x=(x_{i}+x_{i+1})/2$. Detailed structure of each one of the modules enclosed within the regions $(i),\,(ii),$ is described in figure (8) while region $(iii)$ is described by a module similar to the one shown figure (7). In both colormaps the magnitude of flow velocity on the dashed lines drops to $1\%$ of its’ peak value and the parameters which enter (\ref{eq:zeta F}), (\ref{eq:Stream-tanh}) and (\ref{eq:Stream-tanh-1}) are given by $\beta=\gamma=2,\,n=4,\,E=1.$}}
\end{figure}
The resulting zeta potential for this type of
matching is presented in figure (8d). While the values of the stream
function, and zeta potential match precisely along the common
boundary, $x=0$, their corresponding derivatives experience a discontinuity.
The discontinuity in the stream function derivative, $\Delta$ ,
along the common boundary is given by 
\begin{equation}
	\Delta=2\frac{1+\left(1-n\beta x\right)\mathrm{exp}\left(\beta x\right)}{\left(1+\mathrm{exp}\left(\beta x\right)\right)^{n+1}},\label{eq:Discontinuity}
\end{equation}
and could be made arbitrarily small by extending the length of the
domain or by proper choice of the corresponding parameters. For typical
values in our example, $\Delta$ does not increase above $10^{-6}$,
which would be negligible in typical microfluidic systems. 

Figure (9) illustrates the use of superposition and matching for creation
for a micromixer geometry, in which two streams merge in a Y-junction,
and continue to a serpentine channel consisting of six $4\pi/5$ turns.
Figure (9a) shows the stream function, which is an assembly of basic
flow fields matched together along common boundaries. The magnitude
of flow velocity on the dashed lines drops to $1\%$ of its peak value.
Figure (9b) shows the necessary zeta potential distribution, obtained
from (\ref{eq:zeta F-rewritten}). 

\section{Summary and conclusions}

In this work we studied EOF in a Hele-Shaw configuration, with non-uniform
zeta potential distribution. We demonstrated the use of zeta potential
distributions for patterning of complex flow fields, including microchannel
networks.

The flow fields we obtained in this work are depth-averaged fields.
Care should be taken when analyzing the flow of particles in such
fields; for sufficiently high diffusivity of the particle (i.e. low
Peclet number), an ensemble of particles will move with the average
velocity field. However, at high Peclet, individual particles may
follow particular streamlines and the ensemble will exhibit a behavior
substantially different than that of the average. 

In sections IV-V we demonstrated the construction of complex flow
fields by superposition of basic solutions. We note that (\ref{eq: in-plane mean velocity}),
(\ref{eq:Poisson for pressure}) and (\ref{eq:Poisson for the stream function})
are conformal invariant. This follows directly from the fact that
these equations can be rewritten so that each term will contain equal
number of gradient operators (see \cite{Bazant}). This opens a door to employ conformal mapping machinery, to calculate resultant pressure and flow field due to non-uniform zeta potential patterning in the presence of complex geometries with real walls. It is also worth noting that the governing equations (\ref{eq:Poisson for pressure}) and (\ref{eq:Poisson for the stream function}) for the pressure and depth-averaged stream function can be extended to include the effect of slow varying height between the plates. The modified governing equation will be an EOF driven Reynolds equation in which both the viscous terms and the EOF terms depend on the spatially varying gap between the plates. While the governing equation will remain linear, and thus could potentially be resolved analytically, it is beyond the scope of this work.

In practice, surface patterning could be obtained in several ways,
including chemical modification of the surface, or using an array
of surface electrodes. The latter adds complexity, but has the potential
advantage of allowing surface potentials which are much higher, as
well as dynamic modification of the zeta potential distribution.
Further study would be required to characterize the effect of zeta potential
discretization on the resulting flow fields. Flow
patterning by EOF, particularly by dynamic modification of the surface
potential, may serve a powerful tool for manipulating fluids without
mechanical components in a variety of microfluidic applications.

\acknowledgments{This research was supported by the Israel Science Foundation (grants No. 512/12 and 818/13) and FP7 Marie Curie Career Integration (grant No. PCIG09-GA-2011-293576). S.R. is supported in part by a a Technion fellowship from the Lady Davis Foundation.}


\clearpage 
\begin{thebibliography}{9}


\bibitem{Herr} A. E. Herr, J. I. Molho, J. G. Santiago, M. G. Mungal, T. W. Kenny, and M. G. Garguilo, "Electroosmotic Capillary Flow with Nonuniform Zeta Potential," Anal. Chem. \textbf{72}, 1053--1057 (2000).

\bibitem{Stone} H. A. Stone, A. D. Stroock, and A. Ajdari, "Engineering Flows in Small Devices," Annu. Rev. Fluid Mech. \textbf{36}, 381--411 (2004).

\bibitem{AndersonIdol} J. L. Anderson and W. K. Idol, "Electroosmosis Through Pores with Nonuniformly Charged Walls," Chem. Eng. Commun. \textbf{38}, 93--106 (1985).

\bibitem{Ghosal} S. Ghosal, "Lubrication theory for electro-osmotic flow in a microfluidic channel of slowly varying cross-section and wall charge," J. Fluid Mech. \textbf{459}, 103--128 (2002).

\bibitem{Ajdari95} A. Ajdari, "Electro-Osmosis on Inhomogeneously Charged Surfaces," Phys. Rev. Lett. \textbf{75}, 755--758 (1995).

\bibitem{Ajdari96} A. Ajdari, "Generation of transverse fluid currents and forces by an electric field: Electro-osmosis on charge-modulated and undulated surfaces," Phys. Rev. E \textbf{53}, 4996--5005 (1996).

\bibitem{Long} D. Long, H. A. Stone, and A. Ajdari, "Electroosmotic Flows Created by Surface Defects in Capillary Electrophoresis," J. Colloid Interface Sci. \textbf{212}, 338--349 (1999).

\bibitem{Ajdari01} A. Ajdari, "Transverse electrokinetic and microfluidic effects in micropatterned channels: Lubrication analysis for slab geometries," Phys. Rev. E \textbf{65}, 016301 (2001).

\bibitem{Stroock} A. D. Stroock, M. Weck, D. T. Chiu, W. T. S. Huck, P. J. A. Kenis, R. F. Ismagilov, and G. M. Whitesides, "Patterning Electro-osmotic Flow with Patterned Surface Charge," Phys. Rev. Lett. \textbf{84}, 3314--3317 (2000).

\bibitem{Hunter} R. J. Hunter, \textit{Foundations of Colloid Science}, Second Edition (Oxford University Press, 2000).

\bibitem{Dukhin} S. S. Dukhin, "Non-equilibrium electric surface phenomena," Adv. Colloid Interface Sci. \textbf{44}, 1--134 (1993).

\bibitem{Lyklema} J. Lyklema, \textit{Fundamentals of Interface and Colloid Science. Volume 2: Solid-Liquid Interfaces. With Special Contributions by A. de Keizer, B.H. Bijsterbosch, G.J. Fleer and M.A. Cohen Stuart.} (Academic Press, 1995).

\bibitem{Yariv} E. Yariv, "Electro-osmotic flow near a surface charge discontinuity," J. Fluid Mech. \textbf{521}, 181--189 (2004).

\bibitem{KhairSquires} A. S. Khair and T. M. Squires, "Surprising consequences of ion conservation in electro-osmosis over a surface charge discontinuity," J. Fluid Mech. \textbf{615}, 323--334 (2008).

\bibitem{Deryaguin} B. V. Deryaguin, S. S. Dukhin, and A. A. Korotkova, "Diffusiophoresis in electrolyte solutions and its role in the mechanism of film formation from rubber latexes by the method of ionic deposition,'' Kolloidn. Zh. \textbf{23}, 53 (1961).

\bibitem{Prieve} D. C. Prieve, J. L. Anderson, J. P. Ebel, and M. E. Lowell, "Motion of a particle generated by chemical gradients. Part 2. Electrolytes,'' J. Fluid Mech. \textbf{148}, 247 (1984).

\bibitem{KhairSquires2} A. S. Khair and T. M. Squires, "Fundamental aspects of concentration polarization arising from nonuniform electrokinetic transport," Phys. Fluids \textbf{20}, 087102 (2008).

\bibitem{Batchelor} G. K. Batchelor, \textit{An Introduction to Fluid Dynamics}, p. 216-224 (Cambridge University Press, 2000).

\bibitem{HeleShaw} H. S. Hele-Shaw, "Flow of Water," Nature \textbf{58}, 520--520 (1898).

\bibitem{BrothertonDavis} C. M. Brotherton and R. H. Davis, "Electroosmotic flow in channels with step changes in zeta potential and cross section," J. Colloid Interface Sci. \textbf{270}, 242--246 (2004).

\bibitem{Leal} L. G. Leal, \textit{Advanced Transport Phenomena: Fluid Mechanics and Convective Transport Processes} (Cambridge University Press, 2007).

\bibitem{KatzPlotkin} J. Katz and A. Plotkin, \textit{Low-Speed Aerodynamics}, 2 edition (Cambridge University Press, 2001).

\bibitem{Bazant} M. Z. Bazant, "Conformal mapping of some non-harmonic functions in transport theory," Proc. R. Soc. Lond. Math. Phys. Eng. Sci. \textbf{460}, 1433--1452 (2004).


\end{thebibliography}
\end{document}